\documentclass[twocolumn,showpacs,amsmath,amssymb,aps]{revtex4}

\usepackage[dvips]{graphics,color}
\usepackage{latexsym}
\usepackage{graphicx}

\def\dd{{\rm d}}
\def\pd{\partial}
\def\arcsinh{{\,\mbox{arcsinh}}}

\def\eps{\epsilon}

\begin{document}

\title{Brane-world inflation: slow-roll corrections to the spectral index}

\author{Kazuya Koyama, Andrew Mennim and David Wands}
\affiliation{Institute of Cosmology and Gravitation, University of
  Portsmouth, Portsmouth~PO1~2EG, United Kingdom}

\date{September 3, 2007}

\begin{abstract}
We quantify the slow-roll corrections to primordial density
perturbations arising from inflation driven by a four-dimensional
scalar field with a monomial potential in a five-dimensional
non-compact bulk spacetime. Although the difference between the
classical brane-world solutions and standard four-dimensional
solutions is large at early times, the change to the amplitude at
late times of perturbations generated from quantum fluctuations is
first-order in slow-roll parameters, leading to second-order
slow-roll corrections to the spectral index. This confirms that the
leading-order effects are correctly given by previous work in the
literature.
\end{abstract}

\pacs{04.50.+h, 11.10.Kk, 11.10.St \hfill arXiv:0709.0294}

\maketitle

\section{Introduction}

Recent proposals in theoretical physics have suggested that our
four-dimensional Universe could lie on a brane embedded in a
higher-dimensional space-time~\cite{Review}. One of the
most studied is the Randall--Sundrum (RS) model~\cite{RS99} where our
four-dimensional universe is a brane embedded in five-dimensional
anti-de Sitter space-time (AdS$_5$).  Indeed it is possible in this framework
to have only one brane in a non-compact space and this is the scenario
we will consider in this letter.

This brane-world perspective may dramatically change our picture of the early universe.
In this model, the Friedmann equation is modified~\cite{BDEL} to
\begin{equation}
 \label{ModFriedmann}
 H^2 = \frac{\Lambda_4}{3} + \frac{\kappa_4^2}{3}\rho + \frac{\kappa_5^4 \, \rho^2}{36} \,,
\end{equation}
so that the Hubble parameter $H\!\propto\!\rho$ at high
energy-density.
Bulk gravity is strongly coupled to the dynamics on the brane at
high energy,
which changes the evolution of cosmological perturbations \cite{CHKS}.
Thus, it is important to establish whether the evolution of the perturbations
generated during inflation could be significantly different from standard four-dimensional
slow-roll inflation.  In a recent paper~\cite{KMRWH07} we calculated
these effects for the simplest realization of slow-roll inflation on a brane.
The purpose of this letter is to illustrate the effect of these brane-world corrections
on the spectral index of density perturbations created during inflation in some specific models.

The simplest way to realize inflation in the RS model is to have a single scalar
field, the inflaton, confined to the brane~\cite{MWBH99} and only gravity in the bulk.
The amplitude of the resulting scalar curvature perturbations is given by
\begin{equation}
 \langle {\cal R}_c^2 \rangle^{1/2}
 = \left(\frac{H}{\dot{\phi}} \right) \langle \delta \phi^2 \rangle^{1/2},
 \label{curvature0}
\end{equation}
where $\dot{\phi}$ is the time derivative of the inflaton $\phi$ and $\delta \phi$ is the
inflaton fluctuation on a spatially flat hypersurface.
If one makes the assumption that back-reaction due to metric perturbations
in the bulk can be neglected, the quantum expectation value of the inflaton
fluctuations on super-horizon scales in the de Sitter space-time is
$\langle \delta \phi^2 \rangle = H^2 / (4 \pi^2)$.
It has been shown that the curvature perturbation ${\cal R}_{c}$ on comoving
hypersurfaces is constant outside the horizon in this model~\cite{LMSW} and, indeed, independently
of the gravitational theory
so long as energy-momentum is conserved~\cite{WMLL}.
One can then use Eq.~(\ref{curvature0}) as initial conditions to
calculate observables like Cosmic Microwave Background (CMB)
temperature anisotropy. Although the formula~(\ref{curvature0}) for
the curvature perturbation is exactly the same as in the
four-dimensional case, the relation between the Hubble parameter,
$H$, and the scalar field potential, $V$, is different~\cite{MWBH99}
due to the modified Friedmann equation~(\ref{ModFriedmann}). This
changes the prediction of the spectrum of scalar perturbations for a
given potential and observational predictions of brane-world
inflation have been made using this approach~\cite{obs,RL04}.

A crucial assumption is
that back-reaction due to metric perturbations in the bulk can be
neglected.  In the extreme slow-roll limit this is necessarily
correct because the coupling between inflaton fluctuations and
metric perturbations vanishes; however, this is not necessarily the
case when slow-roll corrections are included in the calculation.
Previous work~\cite{PreviousWork} has shown that such bulk effects
can be subtle and interesting (see also \cite{other} for other
approaches). In particular, sub-horizon inflaton fluctuations on a
brane excite an infinite ladder of Kaluza--Klein modes of the bulk
metric perturbations at first order in slow-roll parameters, and a
naive slow-roll expansion breaks down in the high-energy regime once
one takes into account the back-reaction of the bulk metric
perturbations, as confirmed by direct numerical
simulations~\cite{HK}. However, an order-one correction to the
behaviour of inflaton fluctuations on sub-horizon scales does not
necessarily imply that the amplitude of the inflaton perturbations
receives corrections of order one on large scales; one must
consistently quantise the coupled brane inflaton fluctuations and
bulk metric perturbations.
This requires a detailed analysis of the coupled brane-bulk system
\cite{coupled,KMRWH07}.

\section{The bulk-field equation of motion}

The RS model is a brane-world model where the background spacetime
is AdS$_5$ and matter fields are confined to a 3-brane.  We will be
considering the single-brane model first proposed in
Ref.~\cite{RS99}. The brane-world inflation model under
consideration has the action
\begin{multline}
 S=\frac{1}{2\kappa_5^2} \int_{\cal M} d^5x \sqrt{-^{^{(5)}}\!g} \:\Big\{ R-2\Lambda \Big\} +{}
 \\
 \int_{\pd \cal M} d^4x \sqrt{-^{^{(4)}}\!g}
 \left\{ -\lambda + \frac{1}{\kappa_5^2}K -\frac{1}{2} (\pd\phi)^2 - V(\phi)\right\},
\end{multline}
where $K$ is the trace of the extrinsic curvature of the boundary ${\pd \cal M}$,
the AdS curvature scale is defined by $\Lambda = -6 \mu^2$
and the tension $\lambda = 6 \mu /\kappa_5^2$ is tuned to make the brane flat if it has no other energy density.

The brane is our four-dimensional universe and, because we are
interested in studying early-universe inflation, we will make the
slow-roll approximation, where the induced metric is assumed to be
de Sitter but with a slowly changing Hubble parameter.  We will
describe this in coordinates where the background line element has
the form
\begin{equation}
\label{lineelement}
\dd s^2=N(z)^2\Big[\dd z^2+\frac1{H^2\tau^2}(\dd \vec{x}^2-\dd\tau^2)\Big]\,,
\end{equation}
with the function $N(z)$ given by $N(z)=H / \mu \sinh (Hz)$~\cite{BDEL}.
Here, $H$ is the Hubble parameter, which is constant in a pure de Sitter universe, and the coordinate $\tau$ is conformal time on the brane.  The brane is located at
$z_{\rm b}=H^{-1}\arcsinh(H/\mu)$,
which would be exactly constant for a pure de Sitter universe.
During inflation the universe is approximately de Sitter in that the Hubble parameter $H$ changes slowly as the field rolls slowly down the potential.  This slow-rolling behaviour is usually expressed in terms of slow-roll parameters defined as
\begin{equation}
 \eps=-\frac{\dot{H}}{H^2} \;, \qquad \eta=-\frac{\ddot{\phi}}{H\dot\phi} \;,
 \label{slowrolldef}
\end{equation}
where the overdot denotes derivatives with respect to proper time
on
the brane, not the conformal time $\tau$.
We will use the convention that $\tau$ is positive, so that early
times corresponds to $\tau \to \infty$ and late times to $\tau \to
0$.

The equations of motion for the scalar metric perturbations are most
conveniently expressed in terms of a master variable, a technique
first used in the study of brane-world perturbations in
Ref.~\cite{Mukohyama,KIS}. We define a master variable $\chi$, whose
equation of motion
is~\cite{KMRWH07}
\begin{equation}
\label{eom:chi}
\chi_{,\tau \tau}+k^2\chi-\frac{2}{\tau^2}\chi+\frac{1}{H^2\tau^2}\big(-\chi''+U(z)\chi\big) = 0\,,
\end{equation}
where
\begin{equation}
 U(z)=\frac{H^2}{4}\left( 9 -\frac{1}{\sinh^2(Hz)} \right) \,.
\end{equation}
Note that Eq.~(\ref{eom:chi}) is separable.
The other equations are simplified by defining a new variable, $\xi$, describing the scalar field and metric perturbations on the brane, and given by
\begin{equation}
 \xi = -\left( \frac{d}{d\tau} +\frac{1}{\tau} \right) \left( \frac{\phi_{,\tau}}{H} {\cal R}_c \right)
 -\frac{\dot\phi}{3H\tau} \chi_{\rm b} \,,
 \label{rc}
\end{equation}
in terms of the field $\chi$ evaluated on the brane and the curvature perturbation ${\cal R}_c$ on comoving hypersurfaces.
This new variable obeys the following equation of motion~\cite{KMRWH07}
\begin{equation}
\xi_{,\tau\tau}+k^2\xi = -\frac{k^2\dot\phi}{3H\tau} \chi_{\rm b} \,,
\end{equation}
and the junction condition at the brane is
\begin{equation}
\label{junction}
\left[ \chi' + \left(\frac{1}{2}\frac{N'}{N}+ \frac{\kappa_5^2 \dot{\phi}^2}{3}
\right) \chi \right]_{_{\rm b}}
= \kappa_5^2 \dot\phi H \tau \xi \,.
\end{equation}
Here we have neglected all terms suppressed by the slow-roll
parameters except for the terms responsible for the coupling
between $\xi$ and $\chi$, i.e., the terms that induce the mixing between
the brane and bulk modes.
It is useful to define the time-independent Wronskian of two states
$f=(\xi, \chi)$ and $f^*=(\xi^*, \chi^*)$ by~\cite{KMRWH07}
\begin{equation}
 W(f^*,f)=\xi^* \xi_{,\tau}-\xi \xi_{,\tau}^*+\frac{1}{3\kappa_5^2} \int^\infty_{z_{_{\rm b}}} dz\, \big(\chi^* \chi_{,\tau}
 - \chi \chi_{,\tau}^*\big) \,.
\label{wronskian}
\end{equation}
which is conserved.

At late times, $k \tau \to 0$, a growing bound state solution is~\cite{KMRWH07}
\begin{eqnarray}
\label{boundg}
 \xi_g &=& A_g \, k\tau J_{-1/2}(k\tau) \,, \\
 \chi_g &=& C_g \,\tau \sqrt{\sinh(Hz)} \, \log \left(\frac{\cosh(Hz)-1}{\cosh(Hz)+1}\right)\,, \nonumber
\end{eqnarray}
where $A_g$ is the overall amplitude, while $C_g$ can be determined through $A_g$.
A decaying bound state is given by~\cite{KMRWH07}
\begin{eqnarray}
\label{boundd}
 \xi_d &=& A_d \, k\tau J_{1/2}(k\tau) \,, \\
 \chi_d &=& C_d \,\tau^2 \sqrt{\sinh(Hz)}
 \Bigg[ 1+ \nonumber \\
 && \qquad \frac{\cosh(Hz)}{2} \log \left(\frac{\cosh(Hz)-1}{\cosh(Hz)+1}\right)\Bigg]\,, \nonumber
\end{eqnarray}
where $C_d$ is linearly related to $A_d$.

At early times
(and high energies), $k\tau\to\infty$,
the growing and decaying solutions both take the form~\cite{KMRWH07}
\begin{eqnarray}
\chi &\propto& \tau^{1/3} \exp \Bigg\{ i\tau-i \left(\frac{3\epsilon^2\tau}{2}\right)^{1/3}
 -\frac{z-z_{\rm b}}{2} \nonumber \\
 &&{}-\left(\frac{2\epsilon\tau^2}{3}\right)^{1/3} (z-z_{\rm b})
  +\frac{i}{3} \tau (z-z_{\rm b})^2 \Bigg\} +\text{c.c.} \nonumber \\
\xi &\propto& \exp
\left(i \tau -i\left(\frac{3\epsilon^2\tau}{2}\right)^{1/3} \right) + \text{c.c.} \;.
\label{earlytimesolution}
\end{eqnarray}
The growing and decaying solutions have a different amplitude and phase,
with the phase difference being approximately equal to $\pi/2$.
We see that the amplitude of $\chi(z_b,\tau)$ increases
as $\chi(z_b, \tau)  \propto \tau^{1/3}$ while the amplitude of
$\xi$ tends to a constant at large $\tau$, becoming increasingly dominated
by higher modes and more closely bound to the brane.

The early-time solution (\ref{earlytimesolution}) differs from the solution
for the scalar field in 4D at leading order, where $\xi$ satsifies the equation of motion
\begin{equation}
\xi_{,\tau\tau} + k^2\xi - \frac{2\eps-\eta}{\tau^2}\xi = 0 \,,
\end{equation}
and has solutions at late times given by
\begin{equation}
 \label{late4Dxi}
\xi_g = \frac{2^\nu}{\Gamma(-\nu+1)}(k\tau)^{1/2-\nu} \,,\quad
\xi_d = \frac{2^{-\nu}}{\Gamma(\nu+1)}(k\tau)^{1/2+\nu} \,,
\end{equation}
where $\nu=1/2+2 \epsilon - \eta$,
and at early times by
\begin{equation}
\xi_g = N_d \cos (k\tau + \Delta_g) \,,\quad
\xi_d = N_g \cos (k\tau + \Delta_d)
\label{xi4d}
\end{equation}
where $N_d=N_g$ and $|\Delta_g - \Delta_d| = \nu\pi$ for standard 4D inflation.

One might expect that the ${\cal O}(1)$ difference between the
early-time solutions would lead to a significant difference to the
amplitude of quantum perturbations generated during inflation.  In
fact this does not happen.  We now illustrate how the quantum
effects can be calculated and use the result to determine the
slow-roll corrections to the spectral indices of scalar
perturbations for some monomial potentials.

\section{Summary of Wronskian method}

In Ref.~\cite{KMRWH07} we calculated the growing and decaying mode solutions;
these were found to differ significantly from the solution in the standard 4D inflationary paradigm.
However, the amplitude of large-scale perturbations from the quantum vacuum state in the early universe is only affected at the first order in the slow-roll parameters.  We follow the method developed in Refs.~\cite{wronskian} to relate the amplitude of the growing mode solution to the creation and
annihilation operators, which is summarized here.

In principle, this method involves the construction
of properly normalised positive- and negative-frequency solutions at early
times and the evaluation of the inner product (Wronskian) between these
solutions and the bound states. However, there is a short-cut: all we need
are the values of the real coefficients of the positive and negative frequency parts.
We first explain this short-cut by making use of the familiar four-dimensional
example, and then proceed to the brane-world model.

Let us begin with the standard four-dimensional case.
The Wronskian is defined as
\begin{equation}
W(\xi^*,\xi) = \frac{1}{k^2} (\xi^* \xi_{,\tau} - \xi \xi^*_{,\tau}).
\end{equation}
We are interested in the quantum field that behaves at late times as
$\hat{\xi} \to \hat{Z} \xi_g$,
where $\hat{Z}$ is a time-independent quantum
operator. This field is quantised on small scales where we can expand
$\hat{\xi}$ in terms of negative- and positive-frequency modes,
\begin{equation}
\hat{\xi} = \hat{a} \varphi^{(-)} + \hat{a}^{\dagger} \varphi^{(+)},
\end{equation}
where $\hat{a}$ and $\hat{a}^{\dagger}$ are annihilation and
creation operators, respectively, which define the vacuum $\hat{a} |0 \rangle =0$.
Note that
because $\tau$ decreases as proper time increases,
the negative- and positive-frequency functions are $\varphi^{(-)}
\propto e^{ik\tau}$ and $\varphi^{(+)} \propto e^{-i k \tau}$. The
mode functions should be normalised as
$W(\varphi^{(-)},\varphi^{(+)}) = -i$ to ensure the canonical
commutational relation between
$\hat\xi$
and its conjugate momentum.

We use the constancy of the Wronskian to express $\hat{Z}$ in terms
of $\hat{a}$ and $\hat{a}^{\dagger}$. At late times, $\hat{Z} =
W(\hat{\xi},\xi_d) / W(\xi_g,\xi_d) $ where the numerator can be
calculated using the early-time solutions and the denominator using
the late-time solutions (\ref{late4Dxi}) as $ W(\xi_g, \xi_d) =
2 / (k\pi)$ up to the first order in slow-roll parameters.
We expand the growing mode and the decaying mode solutions at early
times by
\begin{equation}
 \xi_g = c_g \varphi^{(-)} + c_g^{*} \varphi^{(+)} \,,\quad
 \xi_d = c_d \varphi^{(-)} + c_d^{*} \varphi^{(+)} \,.
 \label{xisplit}
\end{equation}
This allows us to evaluate the expectation value of the square of the operator $\hat{Z}$ as
\begin{equation}
\langle\hat{Z}^{\dagger} \hat{Z} \rangle = k^2 \left(\frac{\pi}{2} \right)^2
|c_d|^2.
\label{fin4d}
\end{equation}
which fully characterizes the Gaussian random field.
Thus, the problem reduces to finding $c_d$, the expansion coefficient
for the decaying mode solution.

The positive- and negative-frequency modes have the form
$\varphi^{(\pm)} = |\varphi| e^{\mp i \delta} e^{\mp i k \tau}$,
where the normalisation factor $|\varphi|$ is not needed here, and $\delta$ is some phase (which is
irrelevant in the four-dimensional case).
Comparing the two expressions (\ref{xi4d}) and (\ref{xisplit}), and using the Wronskian $W(\xi_g, \xi_d)$,
$|c_d|^2$ is determined to be
\begin{equation}
|c_d|^2 = \frac{1}{2} \frac{N_d}{N_g} \frac{1}{\sin |\Delta_g-\Delta_d|}
\frac{1}{k} \left(\frac{2}{\pi} \right)\,.
\label{cd-4d}
\end{equation}
Thus, the quantity $|c_d|^2$ entering (\ref{fin4d}) is expressed in terms of
the ratio $N_d/N_g$ of the amplitudes and difference $(\Delta_g - \Delta_d)$
of the phases of the decaying and growing modes evolved back in time.
These are evaluated analytically as
$N_d /N_g=1$ and $|\Delta_d -\Delta_g| = \pi/2$, giving
\begin{equation}
|c_d|^2 = \frac{1}{k\pi} \,, \quad
\langle \hat{Z}^{\dagger}\hat{Z} \rangle = \frac{k \pi}{4}.
\label{zz}
\end{equation}
The quantity that is related to observables is the comoving
curvature perturbation ${\cal R}_c$, which is defined as
\begin{equation}
\xi = - \left[\frac{d}{d \tau} + (1+2 \eps-\eta) \frac{1}{\tau}
\right] \left( \frac{\phi_{,\tau}}{H} {\cal R}_c \right).
\end{equation}
Thus, one finds the power spectrum of the curvature perturbation to be
\begin{equation}
\label{PowerSpectrum}
 {\cal P}_{{\cal R}_c}^{1/2}(k) = \big[ 1 + C_1\eps + C_2 \eta \big] \left.\frac{H^2}{2\pi\dot\phi}\right|_{k=aH} \,,
\end{equation}
where $C_1 = 3- 2\ln 2 -2\gamma$ and $C_2 = -2 + \ln 2+\gamma$,
using the fact that conformal time is
$\tau = (1+\epsilon)/ (aH)$ up to the first order
in slow-roll parameters.

The reason for preferring this method for dealing with the brane-world case is that one does not need to know the properly-normalised positive- and negative-frequency modes at early times, merely the quantities $N_d/N_g$ and $|\Delta_d-\Delta_g|$.
For the brane-world a similar calculation presented in our previous work~\cite{KMRWH07} allow us to include the effect of interactions between the brane and bulk.  The Wronskian method, using the formula in Eq.~(\ref{wronskian}), allows us to deal with complicated early-time behaviour.
The expectation value of the square of the operator $\hat{Z}$ is obtained as
\begin{equation}
\langle \hat{Z}^{\dagger}\hat{Z} \rangle = \frac{k \pi}{4} \Big[ 1+K(\eps;H/\mu) \Big] \,,
\end{equation}
where $K(\eps; H/\mu)$ is determined by numerical solution. The left
panel of Fig.~1 shows the dependence of K on the energy scale of the
inflation $H/ \mu$. At low energies,
$H/\mu\ll1$, the correction $K$ is negligible and we recover the
standard vacuum (\ref{zz}).
In the high-energy limit, $H/\mu \gg 1$, the correction can be
approximated as a linear function of $\eps$ , which we write as
$K(\eps;H/\mu) = K_1 \eps$. The right panel of Fig.~1 shows
$K(\eps;H/\mu)/\eps$ in the high energy limit.
This modifies Eq.~(\ref{PowerSpectrum}) for the power spectrum by
changing the coefficient $C_1$ to
\begin{equation}
\tilde{C}_1= 3-2\ln2-2\gamma+(K_1/2) \,, \label{ModifiedC1}
\end{equation}
while $\tilde{C}_2=C_2$.

\begin{figure*}[t]
 \begin{center}
\includegraphics[width=15cm]{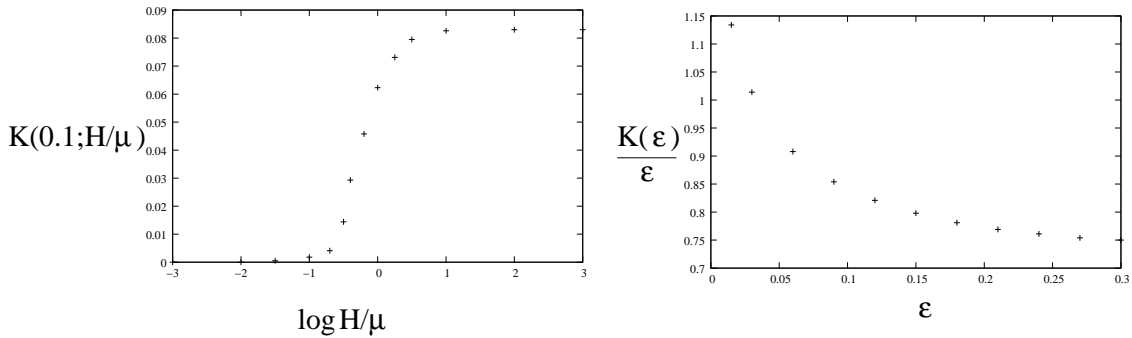}
\caption[]{{\it Left:} $H/\mu$ dependence of $K(\eps;H/ \mu)$ for $\eps=0.1$.
{\it Right:} $K(\eps)/\eps$ in the high energy limits.}
\end{center}
\end{figure*}

\section{Corrections to the spectral index}

One of the strengths of the slow-roll inflation paradigm is that the
perturbations generated by quantum fluctuations have a power
spectrum that is almost scale-invariant.
In 4D inflation the first-order correction to the tilt, $n-1 \equiv
\dd \ln {\cal P}_{{\cal R}_c} / \dd \ln k$, is well known to be $n-1 = -4\eps
+2\eta$. The second-order correction was calculated by Stewart
and Lyth~\cite{SL93}. This was extended to high energy inflation in
the RS model in Ref.~\cite{Cal}, including the modified effect of
the Friedmann equation (\ref{ModFriedmann}) in
Eq.~(\ref{PowerSpectrum}) but without including bulk gravity effects
on the initial vacuum state. This gives
\begin{eqnarray}
\label{StewartLyth}
 n-1 &+& 4\eps - 2\eta = 2(C_1-2) \eps^2 \nonumber\\
 &&+(-4 C_1+2 C_2+2) \eps \eta
 - 2 C_2 \delta^2
   \,,
\end{eqnarray}
where $\delta^2 =\dddot{\phi}/(H^2 \dot{\phi})-\eta^2$.
The actual values of $\eps$ and $\eta$ for any given potential will
change.  Alternative slow-roll parameters $\eps_{_V}$ and
$\eta_{_V}$ are defined in terms of the potential by
\begin{equation}
 \eps_{_V} = \frac{2\lambda}{\kappa_4^2} \frac{{V'}^2}{V^3} \,, \qquad
 \eta_{_V} = \frac{2\lambda}{\kappa_4^2} \frac{V''}{V^2} \,,
\end{equation}
for brane-world inflation in the high-energy limit and
\begin{equation}
 \eps_{_V} = \frac{1}{2\kappa_4^2} \frac{{V'}^2}{V^2} \,, \qquad
 \eta_{_V} = \frac{1}{\kappa_4^2} \frac{V''}{V} \,,
\end{equation}
for standard 4D inflation.
These are related to the slow-roll parameters given in Eq.(\ref{slowrolldef}) by $\eps \approx \eps_{_V}$ and $\eta \approx \eta_{_V} - \eps_{_V}$.
Note that the different Friedmann relation means that the field, $\phi$ will have a different value at horizon crossing.

The calculations
presented in the preceding section
show that we can determine the spectral index to second order for
brane-worlds, including bulk gravity effects at high energies as
well as the modified Friedmann equation, by using the formula given
in Eq.~(\ref{PowerSpectrum}), replacing $C_1$ with $\tilde{C}_1$ as
given in Eq.~(\ref{ModifiedC1}).

\section{Results for Monomial Potentials}

Tables~\ref{Table:MonomialPotentials1}
and~\ref{Table:MonomialPotentials2} show the slow-roll parameters
and spectral indices for some monomial potentials. The differences
at first order arise from the modified evolution on the brane, not
from the brane-bulk mixing.  The second-order corrections in
brane-worlds have a contribution from the brane-bulk
mixing~\cite{KMRWH07} in addition to the familiar Stewart--Lyth
correction, and these are of the same order of magnitude.  For both
the standard and brane-world inflation models the second-order
correction is less than 1 part in $10^3$ so the first-order result
should be sufficient for observational cosmology.
\begin{table}
\begin{tabular}{|c|cc|ccc|}
\hline
&\multicolumn{2}{c|}{~4D Inflation~~} & \multicolumn{3}{c|}{High-Energy BW} \\
$\alpha$ & ~~$\eps_{_{V,50}}$ & $\eta_{_{V,50}}$ & ~~$\eps_{_{V,50}}$ & $\eta_{_{V,50}}$ & $K_1$\\
\hline
2 & $\frac{1}{101}$ & $\frac{1}{101}$ & $\frac{1}{101}$ & $\frac{1}{202}$ & $1.21$ \\
4 & $\frac{1}{51}$ & $\frac{3}{102}$  & $\frac{1}{76}$ & $\frac{3}{304}$ & $1.15$ \\
6 & $\frac{3}{103}$ & $\frac{5}{103}$ & $\frac{3}{203}$ & $\frac{5}{406}$ & $1.14$ \\
\hline
\end{tabular}
\caption{Showing the values of the slow-roll parameters $\eps_{_V}$ and $\eta_{_V}$ for three monomial potentials $V \propto \phi^\alpha$ and 50 e-folds of inflation.}
\label{Table:MonomialPotentials1}
\end{table}

\begin{table}
\begin{tabular}{|c|c@{~~}c|c@{~~}c|}
\hline
&\multicolumn{2}{c|}{~4D Inflation~~} & \multicolumn{2}{c|}{High-Energy BW} \\
$\alpha$ & $1^{\rm st}\!$ order & $2^{\rm nd}\!$ order & $1^{\rm st}\!$ order & $2^{\rm nd}\!$ order \\
\hline
2 & 0.9604 & 0.9601 & 0.9505 & 0.9507 \\
4 & 0.9412 & 0.9389 & 0.9408 & 0.9406 \\
6 & 0.9223 & 0.9169 & 0.9360 & 0.9355 \\
\hline
\end{tabular}
\caption{Showing the spectral index for scalar perturbations with 50 e-folds of inflation to first and second order in the slow-roll parameters.}
\label{Table:MonomialPotentials2}
\end{table}

In the standard inflation paradigm slow-roll corrections change the
dynamics on sub-horizon scales by introducing an effective mass that
is negligible at early times (though this is not negligible on
super-horizon scales).  The problem thus reduces to quantizing a
massless scalar field in de Sitter space. By contrast, in the
brane-world model it is essential to use the slow-roll approximation
even to formulate the problem in a way where the bulk wave equation
is separable. In the brane-world model we have shown that the
coupling to bulk metric perturbations cannot be ignored in the
equations of motion.  Indeed, we have shown that there are
order-unity differences between the classical solutions without
coupling and with slow-roll induced coupling.  However, the change
in the amplitude of quantum-generated perturbations is at
next-to-leading order~\cite{KMRWH07} because there is still no
mixing at leading order between positive and negative frequencies
when scales observable today crossed the horizon, so the Bogoliubov
coefficients receive no corrections at leading order.  The amplitude
of perturbations generated is also subject to the usual slow-roll
corrections on super-horizon scales.  Thus, the correction we
calculate is in addition to the usual Stewart--Lyth
correction~\cite{SL93} and we find it to be of the same order of
magnitude.

Our results also show that the
ratio of tensor-to-scalar perturbation amplitudes are not influenced
by brane-bulk interactions at leading order in slow-roll.  It is not
possible for us to calculate the next-order corrections without also
deriving the slow-roll correction to tensor amplitudes.  This
correction has not yet been calculated because it requires one to go
beyond the approximation of a separable
wave equation in the bulk for a de Sitter brane.

From an end-user's perspective our results are significant in
establishing rigorously the validity of assumptions made in earlier
brane-world literature.  The most significant difference between the
prediction for the spectral indices in the brane-world at high
energy and in standard inflation with the same potential arises at
first order in slow-roll from the modified evolution of the Hubble
parameter and scalar field, leading to different values of the
slow-roll parameters when the number of e-foldings of inflation is
fixed~\cite{MWBH99,obs,RL04}.  It is remarkable that the predictions
from inflation theories should be so robust that this result hold in
spite of the leading-order change to the solutions of the classical
equations of motion.

\section*{Acknowledgments}
The authors are supported by STFC.



\end{document}